\documentclass{jpsj-suppl}
\usepackage{txfonts} 
\usepackage{graphicx,amsmath,amssymb,dcolumn,bm}
\usepackage{slashed}

\title{Useful relations and sum rules for PDFs 
       and multiparton distribution functions of spin-1 hadrons}

\author{S. \textsc{Kumano}$^{1,2}$ and Qin-Tao  \textsc{Song}$^{3,4}$}

\inst{$^{1}$KEK Theory Center,
             Institute of Particle and Nuclear Studies, 
             High Energy Accelerator Research Organization (KEK),
             Oho 1-1, Tsukuba, Ibaraki, 305-0801, Japan \\
$^{2}$ J-PARC Branch, KEK Theory Center,
             Institute of Particle and Nuclear Studies, KEK, 
           and Theory Group, Particle and Nuclear Physics Division, 
           J-PARC Center,  Shirakata 203-1, Tokai, Ibaraki, 319-1106, Japan\\
$^{3}$School of Physics and Microelectronics, Zhengzhou University, 
             Zhengzhou, Henan 450001, China\\
$^{4}$  CPHT, CNRS, Ecole Polytechnique, Institut Polytechnique de Paris,
Route de Saclay, 91128 Palaiseau, France }

\email{songqintao@zzu.edu.cn}

\recdate{January 14, 2022}

\abst{There are two types of polarizations in spin-1 hadrons, 
and they are vector and tensor polarizations.
The latter is a unique one since it does not exist in the spin-1/2 proton.
The vector-polarized PDFs are the same for both the proton and spin-1 hadrons;
therefore, we mainly investigate the unique PDFs in tensor-polarized hadrons. 
By using the operator product expansion, the twist-3 PDF $f_{LT}$ 
can be expressed by two terms in the same way with $g_T$ of the proton. 
The first term is determined by the twist-2 PDF $f_{1LL}$ (or $b_1$)
which was measured by an experiment, and the second term is 
expressed by twist-3 quark-gluon distributions. 
If we neglect the higher-twist effects, $f_{LT}$ is simply 
given by $f_{1LL}$, and this relation is similar 
to the Wandzura-Wilczek  relation of $g_T$.
Furthermore, a new sum rule is also obtained 
for $f_{2LT}=2/3 f_{LT}-  f_{1LL}$, which is analogous to
the  Burkhardt-Cottingham sum rule, in the tensor-polarized spin-1 hadrons. 
In future, these interesting relations could be studied 
at accelerator facilities, such as the Jefferson Laboratory, 
the Fermilab, the nuclotron-based ion collider facility in Russia, 
the proposed electron-ion colliders in US and China.}

\kword{parton distribution, tensor polarization, higher twist effects, sum rule}

\begin{document}
\maketitle

\section{Introduction}
\vspace{-0.05cm}

In 1980s, it was found that the proton spin cannot be explained
only by the combination of quark spins,
and this is known as the proton spin puzzle. 
In order to solve this spin puzzle, one needs to figure out
the contributions from all of the parton's helicities 
and orbital angular momenta.
Generalized parton distributions (GPDs) are investigated 
to obtain the contribution of the orbital angular momenta, 
so that the study of the GPDs has been a hot research topic 
in hadron physics for the past decades.
As for the contribution of the partons' helicities, they can be determined 
by the helicity distributions. At present, the quark helicity distributions 
are relatively well known from many experimental measurements, 
such as by polarized proton-proton collisions 
and deep inelastic scattering (DIS) processes. 
However, the gluon helicity distribution is not accurately obtained, 
and the uncertainty band is still large. 
The  measurement of gluon helicity distribution is one of 
the main physics goals of the proposed Electron-Ion Colliders 
in US and China.
On the other hand, the timelike GPDs were investigated 
by two-photon processes at KEKB \cite{timelike-GPDs}.

Aside from the important helicity distributions in spin physics, 
there are also interesting theoretical studies which
provide the constraints for polarized distributions.
For example, the twist-3 distribution $g_T$ can be decomposed 
into two parts \cite{Jaffe:1991ra}. The first part is determined 
by the helicity distributions, which is known as the Wandzura-Wilczek (WW) 
relation \cite{ww-1977}, and second part is expressed by twist-3 
quark-gluon distributions.
Since the helicity distributions are leading twist PDFs, 
one expects that the WW relation is the main contribution to $g_T$. 
In fact, a recent study confirmed it by showing that 
the WW relation contributes to 60-85\% of $g_T$ in the proton 
\cite{accardi-2009}. 
If we neglect the higher-twist effects in $g_T$, 
one obtains a sum rule for $g_2=g_T-g_1$ as $\int dx g_2(x)=0$, 
and it is called the Burkhardt-Cottingham (BC) sum rule \cite{bc-1970}.

The helicity distributions are related to the vector polarization
in the proton. There is another type, tensor polarization, 
in addition to the vector polarization in spin-1 hadrons. 
One may wonder whether there exist similar relations 
to the WW relation and the BC sum rule for the tensor-polarized PDFs 
in the spin-1 hadrons. 
The spin structure is the same between the proton 
and the vector-polarized spin-1 hadrons; therefore, 
the vector polarization is not discussed in this paper. 
The leading-twist PDF $f_{1LL}$ (or $b_1$) and higher-twist ones 
were investigated in 
Refs.\,\cite{fs83,Hoodbhoy:1988am,ma-spin-1-2013,kumano:2021}.
As for transverse-momentum-dependent parton distribution functions (TMDs)
in the spin-1 hadrons, the leading-twist ones were investigated in Ref.\cite{bm-2000}. 
Recently, a complete decomposition of quark-antiquark correlation function 
was obtained by including the dependence of 
a lightcone vector in the gauge link \cite{kumano:2021}. 
From this correlation function, the TMDs were shown up to the twist 4 
for the tensor-polarized spin-1 hadrons.
Since the complete PDFs and TMDs have been proposed for the spin-1 hadrons, 
it is now possible to study relations among these distributions.
In this work, we show that the twist-3 PDF $f_{LT}$ can be decomposed 
into the contribution of the twist-2 PDF $f_{1LL}$ and 
the one of quark-gluon distributions in the spin-1 hadrons,
in the similar way with $g_T$ \cite{sk:2021}.
Then, a WW-type relation is obtained for $f_{LT}$ and $f_{1LL}$ 
by neglecting the higher twist effects, 
and a BC-type sum rule is shown which provides the constraints for $f_{LT}$.


\section{Parton distributions for tensor-polarized spin-1 hadrons}
\vspace{-0.05cm}

The spin density matrix of spin-1 hadrons is expressed 
by the spin vector $S^{\mu}$ and spin tensor $T^{\mu\nu}$,
and the spin tensor is parameterized as \cite{bm-2000}
\begin{align}
T^{\mu\nu}  = \frac{1}{2} & \left [ \frac{4}{3} S_{LL} \frac{(P^+)^2}{M^2} 
               \bar n^\mu \bar n^\nu 
          - \frac{2}{3} S_{LL} ( \bar n^{\{ \mu} n^{\nu \}} -g_T^{\mu\nu} )
+ \frac{1}{3} S_{LL} \frac{M^2}{(P^+)^2}n^\mu n^\nu
\right.
\nonumber \\
& \ 
\left.
+ \frac{P^+}{M} \bar n^{\{ \mu} S_{LT}^{\nu \}}
- \frac{M}{2 P^+} n^{\{ \mu} S_{LT}^{\nu \}}
+ S_{TT}^{\mu\nu} \right ],
\label{eqn:tensor-1}
\\[-0.90cm] \nonumber
\end{align}
where $n$ and $\bar n$ are lightcone vectors and $P^{\mu}$ 
is the hadron momentum, and they are given by
\begin{align}
n^\mu =\frac{1}{\sqrt{2}} (\, 1,\, 0,\, 0,\,  -1 \, ), \ \ 
\bar n^\mu =\frac{1}{\sqrt{2}} (\, 1,\, 0,\, 0,\,  1 \, ), \ \
P^{\mu}=P^{+} \bar{n}^{\mu}+ \frac{M^2}{2 (P^+)^2} n^{\mu}.
\label{eqn:n-nbar}
\end{align} 
The $S_{LL}$, $S_{LT}^{\mu}$ and $ S_{TT}^{\mu\nu}$ 
in Eq. (\ref{eqn:tensor-1}) are the parameters which 
indicate polarizations of a hadron.
There are two PDFs ($f_{1LL}$, $f_{LT}$)
which are related to vector current 
$  \bar\psi (0) \gamma^\mu \psi (\xi) $ in the lightcone limit of $\xi$, 
and they are defined as 
\cite{Hoodbhoy:1988am,ma-spin-1-2013,kumano:2021} 
\begin{align} 
\langle \, P , T \left | \, \bar\psi (0) 
\, \gamma^\mu \, 
\psi (\xi)  \, \right | \! P, \,  T \,
\rangle _{\xi^+ =0, \, \vec\xi_T=0}  
= \int_{-1}^1 dx e^{-ixP^+ \xi^-}
2 P^+ \left [ \, S_{LL} \, \bar n^\mu \, f_{1LL} (x) 
+ \frac{M}{P^+} \, S_{LT}^{\, \mu} \, f_{LT} (x)  
  \, \right ] .
\label{eqn:vector-matrix}
\end{align}
If one compares Eq.(\ref{eqn:vector-matrix}) with 
the proton's $g_1$ and $g_T$, which are defined by the axial current 
$  \bar\psi (0) \gamma_5 \gamma^\mu \psi (\xi) $, 
they are quite similar. 
The matrix element of Eq.\,(\ref{eqn:vector-matrix}) is written
by considering an arbitrary vector $\xi$ 
\cite{sk:2021}
\begin{align}
 \langle \, P ,  T \left | \,  \bar\psi(0)  \gamma^{\mu}  \psi(\xi) 
  \, \right | P \,  , T \, \rangle
= \int_{-1}^1 dx \, e^{-i x P\cdot \xi}
  \left[ \,  \xi \cdot T \cdot \xi  \left \{  A(x) \, P^\mu
          + B(x) \, \xi^\mu \right \}
          + C(x) \, T^{\mu\nu} \xi_\nu
  \, \right].
\label{eqn:vector-matrix-1}
\end{align}
Here, the Fock-Schwinger gauge $\xi_{\mu}A^{\mu}(\xi) = 0$ is used,
and $A(x)$, $B(x)$, and $C(x)$ are expressed by the PDFs as
\begin{align}
A (x) & = \frac{3 \, M^2}{(P \cdot \xi)^2} \left [ f_{1LL} (x) 
               - \frac{4}{3} f_{LT} (x) \right ], \ 
B (x)  = \frac{3 \, M^4}{2(P \cdot \xi)^3}  \left [ 
               - f_{1LL} (x) + \frac{8}{3} f_{LT} (x)
               \right ] , 
\nonumber \\
C (x) & = \frac{4 \, M^2}{P \cdot \xi} f_{LT} (x)  .
\label{eqn:ABC}
\end{align} 
In the lightcone limit of $\xi$, Eq.(\ref{eqn:vector-matrix-1})  
becomes Eq.(\ref{eqn:vector-matrix}), 
and the Fock-Schwinger gauge turns out to be the lightcone gauge.

\begin{figure}[hp]
\centering
\includegraphics[width=6.0cm]{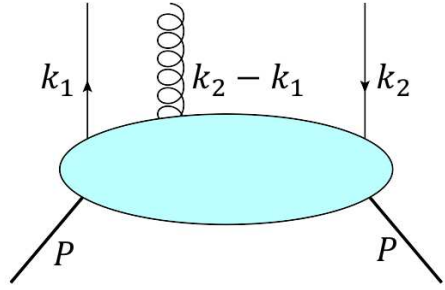} 
\caption{Quark-gluon-quark correlator for a tensor-polarized spin-1 hadron.}
\label{qgq}
\end{figure}

In order to decompose $f_{LT}$ at the twist-3 level, 
the quark-gluon-quark correlator for a tensor-polarized hadron
\begin{align}
(\Phi_G^\alpha)_{ij} (x_1, x_2)=
\int  \! \frac{d \xi_1^-}{2\pi}   \frac{d \xi_2^-}{2\pi}  
 \,    e^{i x_1 P^+ \xi_1^-}  e^{i (x_2-x_1) P^+ \xi_2^-} 
\langle \, P, T \left | \,  \bar\psi _j (0) \, 
g \, G^{+ \alpha}( \xi_2^- ) \, \psi _i (\xi_1^-)  \,
  \right | P, T \, \rangle ,
\label{eqn:3predef-1}
\end{align} 
is needed in the similar way to the $g_T$ decomposition.
This correlator is shown in Fig.\,\ref{qgq}, 
where $k_1^+=x_1 P^+$ and $k_2^+=x_2 P^+$ are the momenta of 
the outgoing quark and the incoming antiquark, respectively. 
At  the twist-3 level, $\Phi_G^\alpha(x_1, x_2)$ can be parameterized as 
\cite{ma-spin-1-2013,sk:2021}
\begin{align}
\Phi_G^\alpha (x_1, x_2) = \frac{M}{2} \bigg [  \, &
i S_{LT}^\alpha  \, F_{G,LT}(x_1, x_2)
- \epsilon_{T}^{\alpha \mu} S_{LT \mu} 
\gamma_5  G_{G,LT}(x_1, x_2)  
\nonumber \\
&+ i \frac{2S_{LL}}{3} \gamma^{\alpha} H_{G,LL}^\perp (x_1, x_2) 
+ i S_{TT}^{\alpha \mu} \gamma_{\mu}  H_{G,TT}(x_1, x_2) \bigg ] 
\slashed{\bar n} .
\label{eqn:3predef1-1}
\end{align} 
There are four quark-gluon distributions, and we will use two of them 
($F_{G,LT}$, $G_{G,LT}$)
on the $S_{LT}^{\mu}$ polarization in decomposing $f_{LT}$.


\section{Twist-2 relation and sum rule by operator product expansion}
\vspace{-0.05cm}

We consider the antisymmetric quark-antiquark operator 
$\bar{\psi}(0) \, (  \vec\partial^{\,\mu} \gamma^{\,\alpha}
  - \vec\partial^{\,\alpha} \gamma^{\,\mu} ) \, \psi (\xi) $
which indicates higher-twist effects, and it is expressed
by quark-gluon-antiquark operators. 
In this calculation, the twist-4 terms and total derivate terms are neglected, 
and we obtain
 \begin{align}
\xi_{\mu} \, \bar{\psi}(0) \, ( 
    \vec \partial^{\,\mu} \gamma^{\,\alpha}
   - \vec{\partial}^{\,\alpha} \gamma^{\,\mu} ) \, \psi (\xi) 
  = \! g  \! \int^1_0 \! dt \, \bar{\psi}(0) \! \left[  
i(t-\frac{1}{2}) G^{\alpha \mu}(t \xi) 
- \frac{1}{2} \gamma_5 \tilde{G}^{\alpha \mu}(t \xi)  
 \right] \! 
 \xi_{\mu} \slashed{\xi} \psi (\xi),
  \label{eqn:gl12-1}
\end{align}
where $\tilde G^{\mu\nu} = \epsilon^{\mu\nu\rho\sigma} G_{\rho\sigma}/2$ 
is the dual field tensor with the convention $\epsilon^{0123}=1$.
We notice that there is no term which 
is dependent on the quark mass, and this is different from the $g_T$ study 
where the operator
$\bar{\psi}(0) \,  (  \vec{ \partial}^{\,\mu}  \gamma^{\,\alpha} 
 - \vec{\partial}^{\,\alpha} \gamma^{\,\mu} ) \gamma_5 \, \psi (\xi) $ 
is investigated \cite{Balitsky:1987bk,Kodaira:1998jn}. 
The left-hand side of Eq. (\ref{eqn:gl12-1}) is the quark-antiquark operator, 
and its matrix element can be expressed by the PDFs defined in 
Eq.\,(\ref{eqn:vector-matrix-1}). Similarly, the matrix element 
of the right side is determined by the quark-gluon distributions 
in Eq.\,(\ref{eqn:3predef1-1}). We combine these results to obtain
the relation\cite{sk:2021}
\begin{align}
x \, \frac{df_{LT}(x)}{dx} + \frac{3}{2} f_{1LL}(x) = - f_{LT}^{(HT)}(x),
\label{eqn:pdfredef4}
\end{align}
where the higher-twist (HT) term $f_{LT}^{(HT)}(x)$ is given
by the principle integral (${\cal P}$) as
\begin{align}
f_{LT}^{(HT)} (x)
= - {\cal P} \int_{-1}^1 dy \frac{1}{x-y} 
  \bigg[ \,  \frac{\partial}{\partial x} 
   \left \{ F_{G,LT} (x, y) 
+ G_{G,LT} (x, y) \right \}
   + \frac{\partial}{\partial y}  \left \{ F_{G,LT} (y, x) 
+ G_{G,LT} (y, x) \right \} \bigg] .
\label{eqn:fbar-twist-3}
\end{align}
Integrating Eq.\,(\ref{eqn:pdfredef4}) over $x$, we obtain
\begin{align}
f_{LT}(x)= \frac{3}{2} \int^{\epsilon (x)}_x \frac{dy}{y} f_{1LL}(y)
+\int^{\epsilon (x)}_x \frac{dy}{y} f_{LT}^{(HT)}(y),
\label{eqn:ww3}
\end{align}
where $\epsilon (x)=|x|/x$ is the sign function, 
and the functions at minus $x$ indicate 
antiquark distributions.
Equation (\ref{eqn:ww3}) provides a decomposition of $f_{LT}$ 
at the twist-3 level. The first term of the right-hand side is expressed 
by the leading-twist distribution $f_{1LL}$, while the second one 
is determined by the twist-3 quark-gluon distributions.
If we define the plus function as $ f^+ (x) \equiv f (x) + \bar f (x)$ 
to combine the quark and antiquark distributions, we have
\begin{align}
f_{LT}^+(x)= \frac{3}{2} \int^1_x \frac{dy}{y} \, f_{1LL}^+ (y)
+\int^1_x \frac{dy}{y} \, f_{LT}^{(HT)+}(y) ,
\label{eqn:pdfredef5}
\end{align}
from Eq.\,(\ref{eqn:ww3}). Here, the distribution functions are
given in the position $x$ region, and
$ f^+ (x)$ is a single flavor distribution function.
Neglecting the twist-3 contributions, we obtain
\begin{align}
f_{2LT}^+ (x)=-f_{1LL}^+ (x)+ \int^1_x \frac{dy}{y} \, f_{1LL}^+(y), \,   
f_{2LT}(x) \equiv \frac{2}{3} f_{LT}(x) - f_{1LL}(x).
\label{eqn:pdfredef7}
\end{align}
The function $f_{2LT}(x)$ is similar to $g_2(x)$ in the proton, 
and Eq. (\ref{eqn:pdfredef7}) is the counterpart of the WW relation 
in the proton. 
Integrating Eq.\,(\ref{eqn:pdfredef7}) over $x$,  we find
a new sum rule as
\begin{align}
\int_0^1 dx \, f_{2LT}^+ (x) =0.
 \label{eqn:pdfredef8}
\end{align}
In addition, if the parton-model sum rule
$\int dx f_{1LL}^+ (x) = 0$ $\left ( \int dx b_1 (x) = 0 \right )$
\cite{b1-sum} is valid with
vanishing tensor-polarized antiquark distributions,
we have another sum rule for $f_{LT}$ itself as
$ \int_0^1 dx \, f_{LT}^+ (x) =0 $.
Recently, we also derived relations among
the twist-3 PDF $f_{LT}$,
the trasverse-momentum moment PDF $f_{1LT}^{\,(1)}$, and 
the multiparton distribution functions 
and among
the twist-3 PDF $e_{LL}$, the twist-2 PDF $f_{1LL}$,
and a multiparton distribution function
\cite{eq-motion}.
All of these relations are useful for future theoretical and
experimental investigations on the structure functions
of the spin-1 hadrons.


\vspace{0.70cm}
\section{Summary}
\vspace{-0.05cm}

The tensor polarization is available for the spin-1 hadrons in comparison
with the spin-1/2 proton, and there are four PDFs which are associated 
with the tensor polarization. In this work, we mainly investigated the relations 
between $f_{1LL}$ and $f_{LT}$ theoretically.
First, $f_{LT}$ was expressed by the twist-2 contribution and the twist-3 one. 
Second, the Wandzura-Wilczek-type relation was derived by neglecting 
the higher-twist effects in $f_{LT}$. Finally, we obtained 
the sum rule for $f_{2LT}$ which is analogous to the Burkhardt-Cottingham
sum rule for the proton.
For the Jefferson Laboratory and Fermilab experiments, the tensor-polarized 
deuteron target is now under development, and these interesting relations 
could be investigated by the experiments in the near future.


\section*{Acknowledgements}
\vspace{-0.4cm}
S. Kumano was partially supported by 
Japan Society for the Promotion of Science (JSPS) Grants-in-Aid 
for Scientific Research (KAKENHI) Grant Number 19K03830.
Qin-Tao Song was supported by the National Natural Science Foundation 
of China under Grant Number 12005191, the Academic Improvement Project 
of Zhengzhou University, and the China Scholarship Council 
for visiting Ecole Polytechnique.


\end{document}